\documentclass[onecolumn, secnumarabic, amssymb, nobibnotes, aps, showpacs, superscriptaddress,twoside,12pt]{article}
\usepackage{graphicx,subfigure,amsmath}%
\usepackage{color}
\usepackage[misc]{ifsym}
\usepackage{float}
\usepackage{epstopdf}
\usepackage{indentfirst}
\usepackage{bm}
\usepackage{epsfig}
\usepackage{enumerate}
\usepackage{amsthm,amsmath,amssymb}

\usepackage{mathrsfs}
\usepackage{graphicx}

\usepackage{xcolor}

\usepackage[
backend=bibtex,
style=numeric-comp,
sorting=none
]{biblatex}
\begin{document}

%\begin{CJK*}{GBK}{song}

%-------------------  First Head  -----------------------------------------
%\thispagestyle{empty} \vspace*{0.8cm}\hbox
%to\textwidth{\vbox{\hfill\huge\sf Modern Physics Letter B / SPIN\hfill}}
%\par\noindent\rule[3mm]{\textwidth}{0.2pt}\hspace*{-\textwidth}\noindent
%\rule[2.5mm]{\textwidth}{0.2pt}

%=================== Text begin here =============================================

\begin{center}
\LARGE\bf The magnon mediated plasmon friction: a functional integral approach%$^{*}$
\end{center}

%\footnotetext{\hspace*{-.45cm}\footnotesize $^*...........$.}
\footnotetext{\hspace*{-.45cm}\footnotesize $^*$Corresponding author, E-mail: fyliu@cxtc.edu.cn}

\begin{center}
\rm Yang Wang$^{\rm a)}$, \ \ Ruanjing Zhang$^{\rm b)}$, \ and  Feiyi Liu\ $^{\rm c) *}$
\end{center}

\begin{center}
\begin{footnotesize} \sl
School of Big Data and Basic Science, Shandong Institute of Petroleum and Chemical Technology, Dongying, 257061, China$^{\rm a)}$ \\
Institute of Theoretical Physics, School of Science, Henan University of Technology, Zhengzhou, 450001, China$^{\rm b)}$ \\
School of Physics and Electronic Science, Chuxiong Normal University, Chuxiong, 675000, China$^{\rm c)}$ \\
\end{footnotesize}
\end{center}

\begin{center}
%\footnotesize (Received XXXX; revised manuscript received XXXX)
(Date: August 2nd, 2024)
\end{center}

\vspace*{2mm}
\begin{abstract}
%\begin{center}
%\begin{minipage}{15.5cm}
%\parindent 20pt\footnotesize
In this paper, we discuss quantum friction in a system formed by two metallic surfaces separated by a ferromagnetic intermedium of a certain thickness. The internal degrees of freedom in the two metallic surfaces are assumed to be plasmons, while the excitations in the intermediate material are magnons, modeling plasmons coupled to magnons. During relative sliding, one surface moves uniformly parallel to the other, causing friction in the system. By calculating the effective action of the magnons, we can determine the particle production probability, which shows a positive correlation between the probability and the sliding speed. Finally, we derive the frictional force of the system, with both theoretical and numerical results indicating that the friction, like the particle production probability, also has a positive correlation with the speed.
%\end{minipage}
%\end{center}
\end{abstract}

%\begin{center}
%\begin{minipage}{15.5cm}
%\begin{minipage}[t]{2.3cm}{\bf Keywords:}\end{minipage}
%\begin{minipage}[t]{13.1cm}
%Frictional force, magnon, plasmon, functional integral.
%\end{minipage}\par\vglue8pt

%\end{minipage}
%\end{center}

\section{Introduction}
In the study of condensed matter physics, the discussions on frictional forces in microscopic systems are always of interest. In these systems, the quantum fluctuations play a crucial role in the dissipation process of frictional force. Generally speaking, the internal relative motion can excite the internal degrees of freedom (DOFs) of the system through the interaction between these internal DOFs, which leads to the production of quasi-particles by the kinetic energy of the internal relative motion. Accordingly, the energy dissipation processes of the frictional force in different systems are carried out by different elementary excitations, such as phonons, electrons, plasmons, magnons, etc~\cite{Spencer1999,PERSSON199983,Mason2001,PhysRevLett.80.1690}. At high temperatures, the dissipation process is usually dominated by phonons, and the classical dynamics of the phonons make the major contribution.
As the temperature decreases, quantum dynamics gradually come into play, and in the dissipation processes, electrons, plasmons, or other collective excitations take on dominant roles. In the past, the corresponding friction in these processes has been widely studied~\cite{Pendry1997,PhysRevB.72.224101,Kheiri2023,Wangyang2022,PhysRevB.108.045406,Khosravi2024}, and with the development of experimental technology, magnetic friction has also been observed~\cite{PhysRevLett.101.137205}. The dissipation of this friction is carried out by magnons, so magnetic friction is also called spin friction.

As magnetic materials are controlled down to the nanometer scale, the dissipation and friction in quantum spin systems have attracted considerable interest. The occurrence of magnetic friction was studied theoretically using a nanometer-sized tip scanning a magnetic surface, examining the dynamics of a classical spin model interacting through dipolar and exchange interactions~\cite{PhysRevB.77.174426}. And a similar approach was also applied to discuss the temperature-dependence of the magnetic friction~\cite{Magiera5257437}. Kadau studied the magnetic friction between two Ising spin systems and found that near the critical temperature the friction is the strongest~\cite{PhysRevLett.101.137205}. These works all focused on purely classical models.
By considering the contribution of quantum fluctuations, we found the bosonic model to be almost equivalent to the general model of electronic friction, by transforming the Heisenberg model into a bosonic model via the Holstein-Primakoff transformation~\cite{Wang2022wyh}.
The above studies on frictional force are based on models that the interfaces consist of two identical surfaces or one surface coupled to a nanoparticle. In reality, friction usually occurs at interfaces composed of two different materials, and in many cases, quantum spins can lead to friction by interacting with other DOFs.  Johan, Brevik and et al studied the Casimir friction between a magnetic and a dielectric material. In this work, the internal DOFs of the magnet were modeled as a single frequency quantum oscillator to obtain the temperature-dependence of the magnetic friction~\cite{PhysRevA.99.042511}.
However the contribution of dispersive magnons has not been discussed, and additionally, the dissipation and frictional force mediated by a intermedium instead of vacuum gap have not been considered.

Most quantum friction models are sandwich-like, i.e., a vacuum gap sandwiched between two objects~\cite{Pendry1997,Wangyang2022,PhysRevA.99.042511}, and the excitations of the vacuum are always photons, refering to `photon-mediated friction'.
In our study, we focus on the frictional force in the case of two objects separated by some intermedium whose excitations are quasi particles.
Specifically, we build a model that the interface is formed by two metallic surfaces separated by a ferromagnetic intermedium. The internal DOFs of the both metallic surfaces are electrons and the excitations of the intermedium are magnons. The whole system is built as plasmons coupled to magnons, and the structure of the model is similar to the  graphene plasmons form polaritons with the magnons of two-dimensional ferrromagnetic insulators~\cite{Costa2023}.
By relative sliding, one surface moves uniformly parallel to the other causing friction in the system.
Our goal is to study the magnon-mediated plasmon friction, through calculation and analysis of the system's effective action of the magnons, particle production probability and etc.

The remaining content of this paper is structured in the following manner. In Section~\ref{model}, we introduce the model to study frictional force.
Section~\ref{effective} gives the effective action of the magnons in the system. Section~\ref{particle} gives the calculation of the probability of particle production. In Section~\ref{friction}, we discuss the friction force of this system. And Section~\ref{conclusion} is the summary of this work.

\section{The model}\label{model}
\begin{figure}[htbp]
  \centering
  \includegraphics[width=5in]{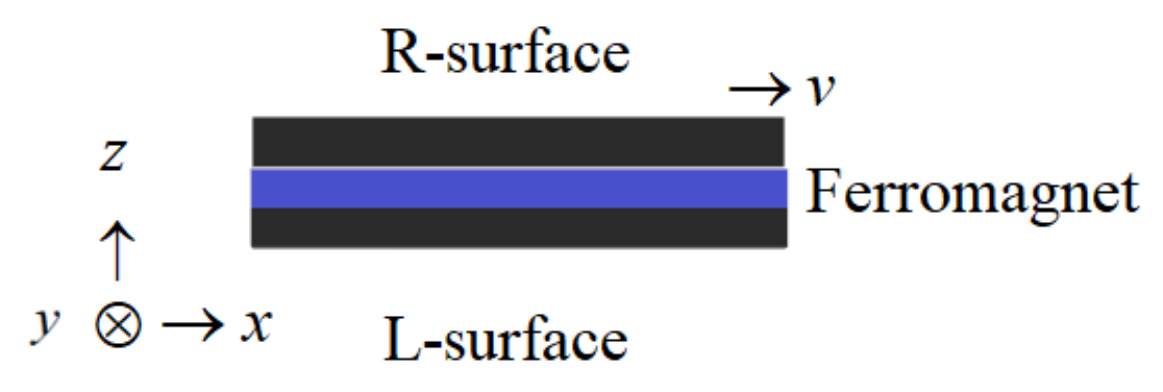}\\
  \caption{The model of interface formed by two metallic surfaces (R- and L-surface in black) separated by a ferromagnetic intermedium (in blue). The R-surface is assumed to slide along the $x-y$ plane with velocity $v$, experiencing a frictional force.}\label{fig_model}
\end{figure}

In this study, we consider the system formed by two metallic surfaces separated by a ferromagnetic intermedium with thickness $h$, and the diagrammatic sketch of the model is shown in Figure~\ref{fig_model}. Due to the metallic nature of both surfaces, the primary basic excitations with spin on the two surfaces are bound to be electrons as the internal DOFs.
The excitations of the intermedium are magnons. The structure of this model is similar to the graphene plasmons form polaritons with the magnons of two-dimensional ferrromagnetic insulators~\cite{Costa2023}. By focusing on the case of plasmons coupled to magnons, our goal is to study the magnon-mediated plasmon friction.
And also, we follow the natural unit such that $\hbar=c=1$.
Thus, the effective internal DOFs of the metallic surfaces that are coupled to the intermedium are assumed to be plasmons, and the Hamiltonian of this plasmons for the two metallic surfaces can be expressed as

\begin{align}
H_\mathrm{R}=\int \mathrm{d}^2\bm{r}_{\parallel}  \left\{  \frac{1}{2}[\pi_\mathrm{R}(\bm{r}_{\parallel})]^2+ \frac{1}{2}[u\nabla\phi_\mathrm{R}(\bm{r}_{\parallel})]^2 + \frac{1}{2}\Omega^2[\phi_\mathrm{R}(\bm{r}_{\parallel})]^2 \right\},
\end{align}
and
\begin{align}
H_\mathrm{L}=\int \mathrm{d}^2\bm{r}_{\parallel}  \left\{  \frac{1}{2}[\pi_\mathrm{L}(\bm{r}_{\parallel})]^2+ \frac{1}{2}[u\nabla\phi_\mathrm{L}(\bm{r}_{\parallel})]^2 + \frac{1}{2}\Omega^2[\phi_\mathrm{L}(\bm{r}_{\parallel})]^2 \right\},
\end{align}
where $\bm{r}_{\parallel}=\begin{pmatrix}x,y\end{pmatrix}$ is the two dimensional Cartesian coordinates in the L-surface. Here we denote the DOFs of the two surfaces via subscripts L and R, i.e., the R-surface and the L-surface.  
The integral measure $\int \mathrm{d}^2\bm{r}_{\parallel} $ means $\int_{-\infty}^{\infty} \mathrm{d}x \int_{-\infty}^{\infty} \mathrm{d}y$. For the  magnons, the Hamiltonian is taken as
\begin{align}
H_\mathrm{M}=\int\mathrm{d}^2\bm{r}_{\parallel} \int_{0}^{h}\mathrm{d}z JS[b^\dagger (\bm{r}_{\parallel},z)\nabla^2 b(\bm{r}_{\parallel},z)],
\end{align}
where $z$ is the third Cartesian coordinate Perpendicular to the L-surface. $J$ is the interchange parameter of the Heinsenberg model, and $S$ is the spin quantum number of the spin magnetic moment in the ferromagnet~\cite{Miura2012}.

In spired by the model of magnon-plasmon coupling in Ref.~\cite{PhysRevB.108.045414}, we define the coupling terms between R/L-surface and the magnet as
\begin{align}
H_\mathrm{RM}=\int \mathrm{d}^2\bm{r}_{\parallel} \int_{-\infty}^{\infty} \mathrm{d}z \delta(z-h) \pi_\mathrm{R}(\bm{r}_{\parallel}) \left[  C(\bm{r}_{\parallel})b(\bm{r}_{\parallel},z) +C^*(\bm{r}_{\parallel})b^\dagger(\bm{r}_{\parallel},z)  \right],
\end{align}
and
\begin{align}
H_\mathrm{LM}=\int \mathrm{d}^2\bm{r}_{\parallel} \int_{-\infty}^{\infty} \mathrm{d}z \delta(z) \pi_\mathrm{L}(\bm{r}_{\parallel}) \left[  C(\bm{r}_{\parallel})b(\bm{r}_{\parallel},z) +C^*(\bm{r}_{\parallel})b^\dagger(\bm{r}_{\parallel},z)  \right].
\end{align}
Here $b(\bm{r}_{\parallel},z)$ and $b^\dagger(\bm{r}_{\parallel},z)$ are the annihilation and creation operators of the Holstein-Primakoff bonsons which describe the magnons~\cite{ERastelli1979}.

The transition amplitude from a ground state to the other is defined as  $Z=\langle 0|U_\mathrm{I}(\frac{T}{2},-\frac{T}{2})|0\rangle$, where  $U_\mathrm{I}(\frac{T}{2},-\frac{T}{2})$ is the time evolution operator of the system and $|0\rangle$ refers to the ground state. This transition amplitude also can be written as a functional integral~\cite{WenXiaoGang2007}:
\begin{align}
Z= \int \mathrm{D}b \int \mathrm{D}b^\dagger \int \mathrm{D}\phi_\mathrm{R} \int \mathrm{D}\phi_\mathrm{L} \mathrm{e}^{\mathrm{i}\int\mathrm{d}^4X (L_\mathrm{R}+L_\mathrm{L}+L_\mathrm{M}+L_\mathrm{RM}+L_\mathrm{LM})  },
\end{align}
where $\int\mathrm{d}^4X=\int_{-\infty}^{\infty}\mathrm{d}t \int_{-\infty}^{\infty}\mathrm{d}x \int_{-\infty}^{\infty}\mathrm{d}y \int_{-\infty}^{\infty}\mathrm{d}z$, integrations over time $t$ and the coordinates. Since the corresponding Lagrangian can be derived using the Legendre transformation, the free parts of the Lagrangian can be written as
\begin{align}
L_\mathrm{R}=\int \mathrm{d}^2\bm{r}_{\parallel} \int_{-\infty}^{\infty}\mathrm{d}z \left\{  \frac{1}{2}\left[\frac{\partial}{\partial t}\phi_\mathrm{R}(t,\bm{r}_{\parallel})\right]^2 - \frac{1}{2}[u\nabla\phi_\mathrm{R}(t,\bm{r}_{\parallel})]^2 - \frac{1}{2}\Omega^2[\phi_\mathrm{R}(t,\bm{r}_{\parallel})]^2 \right\}\delta(z-h),
\end{align}
\begin{align}
L_\mathrm{L}=\int \mathrm{d}^2\bm{r}_{\parallel} \int_{-\infty}^{\infty}\mathrm{d}z \left\{  \frac{1}{2}\left[\frac{\partial}{\partial t}\phi_\mathrm{L}(t,\bm{r}_{\parallel})\right]^2 - \frac{1}{2}[u\nabla\phi_\mathrm{L}(t,\bm{r}_{\parallel})]^2 - \frac{1}{2}\Omega^2[\phi_\mathrm{L}(t,\bm{r}_{\parallel})]^2 \right\}\delta(z),
\end{align}
\begin{align}
L_\mathrm{M}=\int\mathrm{d}^2\bm{r}_{\parallel} \int_{0}^{h}\mathrm{d}z \left[b^\dagger (t,\bm{r}_{\parallel},z)\left(\mathrm{i}\frac{\partial}{\partial t}-JS\nabla^2\right) b(t,\bm{r}_{\parallel},z)\right].
\end{align}
and the interaction parts are expressed as
\begin{align}
L_\mathrm{RM}=-\int \mathrm{d}^2\bm{r}_{\parallel} \int_{-\infty}^{\infty} \mathrm{d}z \delta(z-h) \left[\frac{\partial}{\partial t}\phi_\mathrm{R}(t,\bm{r}_{\parallel})\right] \left[  C(\bm{r}_{\parallel})b(t,\bm{r}_{\parallel},z) +C^*(\bm{r}_{\parallel})b^\dagger(t,\bm{r}_{\parallel},z)  \right],
\end{align}
and
\begin{align}
L_\mathrm{LM}=-\int \mathrm{d}^2\bm{r}_{\parallel} \int_{-\infty}^{\infty} \mathrm{d}z \delta(z) \left[\frac{\partial}{\partial t}\phi_\mathrm{L}(t,\bm{r}_{\parallel})\right] \left[  C(\bm{r}_{\parallel})b(t,\bm{r}_{\parallel},z) +C^*(\bm{r}_{\parallel})b^\dagger(t,\bm{r}_{\parallel},z)  \right].
\end{align}
Here $C(\bm{r}_{\parallel})$ is the interacting strength which is assumed to be $\bm{r}_{\parallel}$ dependent rather than $t$, and $C^*(\bm{r}_{\parallel})$ is the complex conjugation of $C(\bm{r}_{\parallel})$.

\section{The effective action of the magnons}\label{effective}
In our model, the ferromagnet plays a mediating role in the coupling between the plasmons belonging to the R- and L-surface, i.e., the virtual magnons carry the interaction between the two surfaces, which makes it necessary to analyse the free Green's function of the magnons. To achieve this goal, we first derive the effective action of the magnons via a functional integral:
\begin{align}
\mathrm{e}^{\mathrm{i}S_\mathrm{eff}[b,b^\dagger]}= \int \mathrm{D}\phi_\mathrm{R} \int \mathrm{D}\phi_\mathrm{L} \mathrm{e}^{\mathrm{i}\int\mathrm{d}^4X (L_\mathrm{R}+L_\mathrm{L}+L_\mathrm{M}+L_\mathrm{RM}+L_\mathrm{LM})}.
\end{align}
Since it is Gaussian-like shape, the integral can be performed directly and the effective action of the magnon can be expressed as
\begin{align}\nonumber\\&
S_\mathrm{eff}[b,b^\dagger]=\int_{-\infty}^{\infty}\mathrm{d}t \int\mathrm{d}^2\bm{r}_{\parallel} \int_{0}^{h}\mathrm{d}z \left[b^\dagger (t,\bm{r}_{\parallel},z)\left(\mathrm{i}\frac{\partial}{\partial t}-JS\nabla^2\right) b(t,\bm{r}_{\parallel},z)\right]\nonumber\\&
+\int\mathrm{d}^4X \int\mathrm{d}^4X' |C(\bm{r}_{\parallel})|^2 \{b^\dagger (t,\bm{r}_{\parallel},z)\delta(z-h)\Delta_\mathrm{R}(t-t',\bm{r}_{\parallel}-\bm{r}_{\parallel}')\delta(z'-h)\delta(z-h) b(t,\bm{r}_{\parallel},z)\}\nonumber\\&
+\int\mathrm{d}^4X \int\mathrm{d}^4X' |C(\bm{r}_{\parallel})|^2 \{b^\dagger (t,\bm{r}_{\parallel},z)\delta(z)\Delta_\mathrm{L}(t-t',\bm{r}_{\parallel}-\bm{r}_{\parallel}')\delta(z')\delta(z) b(t,\bm{r}_{\parallel},z)\},
\label{Eff_act}
\end{align}
where the kernels are defined as the following Fourier transformations with an infinitesimal positive parameter $\epsilon$:
\begin{align}
\Delta_\mathrm{R}(t-t',\bm{r}_{\parallel}-\bm{r}_{\parallel}')=\int_{-\infty}^{\infty}\frac{\mathrm{d}\omega}{2\pi} \int_{-\infty}^{\infty}\frac{\mathrm{d}k_x}{2\pi} \int_{-\infty}^{\infty}\frac{\mathrm{d}k_y}{2\pi}
\frac{\omega^2\mathrm{e}^{-\mathrm{i}[\omega(t-t')-\bm{k}_{\parallel}\cdot(\bm{r}_{\parallel}-\bm{r}_{\parallel}')]}} {\omega^2-u^2\bm{k}_{\parallel}^2-\Omega^2+\mathrm{i}\epsilon},
\label{Eff_act_ker_R}
\end{align}
and
\begin{align}
\Delta_\mathrm{L}(t-t',\bm{r}_{\parallel}-\bm{r}_{\parallel}')=\int_{-\infty}^{\infty}\frac{\mathrm{d}\omega}{2\pi} \int_{-\infty}^{\infty}\frac{\mathrm{d}k_x}{2\pi} \int_{-\infty}^{\infty}\frac{\mathrm{d}k_y}{2\pi}
\frac{\omega^2\mathrm{e}^{-\mathrm{i}[\omega(t-t')-\bm{k}_{\parallel}\cdot(\bm{r}_{\parallel}-\bm{r}_{\parallel}')]}} {\omega^2-u^2\bm{k}_{\parallel}^2-\Omega^2+\mathrm{i}\epsilon}.
\label{Eff_act_ker_L}
\end{align}
These two kernels can be obtained by taking the second time derivative of the free time-ordered Green's function of the plasmons. Note that the additional frequency factor $\omega^2$ results from the coupling between the canonical momentum of the plasmon and the magnon field. $\bm{k}_{\parallel}=\begin{pmatrix}k_x,k_y\end{pmatrix}$ is the Fourier momentum corresponding to $\bm{r}_{\parallel}$.

Since our purpose is to study frictional force, relative sliding is required in the system. Here we assume that the R-surface is moving uniformly parallel to the L-surface,
and then the terms containing the plasmon DOFs should undergo a transformation, i.e.
\begin{align}
\bm{r}_{\parallel} \rightarrow \tilde{\bm{r}}_{\parallel}=  \bm{r}_{\parallel} - \bm{v}_{\parallel}t,
\end{align}
which makes the Green's function of the R-plasmons transforms as
\begin{align}
\tilde{\Delta}_\mathrm{R}(t-t',\bm{r}_{\parallel}-\bm{r}_{\parallel}')=\int_{-\infty}^{\infty}\frac{\mathrm{d}\omega}{2\pi} \int_{-\infty}^{\infty}\frac{\mathrm{d}k_x}{2\pi} \int_{-\infty}^{\infty}\frac{\mathrm{d}k_y}{2\pi}
\frac{(\omega-\bm{v}_{\parallel}\cdot \bm{k}_{\parallel})^2\mathrm{e}^{-\mathrm{i}[\omega(t-t')-\bm{k}_{\parallel}\cdot(\bm{r}_{\parallel}-\bm{r}_{\parallel}')]}} {(\omega-\bm{v}_{\parallel}\cdot \bm{k}_{\parallel})^2-u^2\bm{k}_{\parallel}^2-\Omega^2+\mathrm{i}\epsilon}.
\end{align}
Then the additional Green's function of the magnons can be calculated via the functional integral from the effective action as follows~\cite{WenXiaoGang2007}:
\begin{align}
    G(X,X')=-\mathrm{i}\frac{\int \mathrm{D}b\mathrm{D}b^\dagger  \mathrm{e}^{\mathrm{i}S_\mathrm{eff}}b(X)b(X')}{\int \mathrm{D}b\mathrm{D}b^\dagger \mathrm{e}^{\mathrm{i}S_\mathrm{eff}}}.
\end{align}

The free Green's function can be obtained by dropping the interaction terms in equation (13).
And in consideration of the boundary conditions of the ferromagnet, this free Green's function of the magnons can be classified into four correlators:

a) the bulk-bulk correlator: $G_\mathrm{F}(t-t',{\bm{r}}_{\parallel}-{\bm{r}}'_{\parallel};z-z')$,

b) the R-surface correlator: $G_\mathrm{F}(t-t',{\bm{r}}_{\parallel}-{\bm{r}}'_{\parallel};z-z'=0)$,

c) the L-surface correlator: $G_\mathrm{F}(t-t',{\bm{r}}_{\parallel}-{\bm{r}}'_{\parallel};z-z'=0)$,

d) the R-L correlator: $G_\mathrm{F}(t-t',{\bm{r}}_{\parallel}-{\bm{r}}'_{\parallel};z-z'=h)$.

\noindent The subscript `F' means Feynmann propagator which corresponds to the free Green's funtion in Quantum Field Theory. Since the coupling between the plasmons and the magnons exists only on the surfaces, it is necessary to calculate the R-L correlator, which can be expressed through Fourier transformation using the standard approach as follows:
\begin{align}
    G_\mathrm{F}(t-t',\bm{r}_{\parallel}-\bm{r}'_{\parallel};0-h)
    =\int\frac{\mathrm{d}\omega}{2\pi}\int\frac{\mathrm{d}^2\bm{k}_{\parallel}}{(2\pi)^2}
     G_\mathrm{F}(\omega,\bm{k}_{\parallel};0-h)\mathrm{e}^{\mathrm{i}[\omega(t-t')-\bm{k}_{\parallel}\cdot(\bm{r}_{\parallel}-\bm{r}'_{\parallel})]},
\end{align}
where the kernel $ G_\mathrm{F}(\omega,\bm{k}_{\parallel};0-h)$ is defined via the inverse Fourier transformation:
\begin{align}
    G_\mathrm{F}(\omega,\bm{k}_{\parallel};0-h):=\int_{-\infty}^{\infty}\frac{\mathrm{d}k_z}{2\pi} \frac{\mathrm{e}^{-\mathrm{i}k_z(0-h)}}{\omega-JS\bm{k}_{\parallel}^2-JSk_z^2+\mathrm{i}\epsilon}=\frac{-\mathrm{ie}^{-\mathrm{i}h\sqrt{\frac{\omega}{JS}-\bm{k}_{\parallel}^2}}}{2\sqrt{\frac{\omega}{JS}-\bm{k}_{\parallel}^2}},
\label{Kel_Four}
\end{align}
and obviously, $G_\mathrm{F}(\omega,\bm{k}_{\parallel};0-h) =  G^*_\mathrm{F}(\omega,\bm{k}_{\parallel};h-0)$.

\section{The probability of particle production}\label{particle}
During the relative sliding, the accumulated energy results from particle production. To obtain the frictional force, we need to track the energy transfer in the system, making it necessary to determine the probability of particle production. Since our model assumes weak interaction, we can calculate the probability of particle production using perturbation theory.

The probability for the system staying at the ground state is $|Z|^2$, so $\mathcal{P}=1-|Z|^2$ means the probability of particle production. 
And $\mathcal{P}$ can be calculated from the effective action of the magnons in Eq \eqref{Eff_act} via functional integral[16]:
\begin{align}
    \mathcal{P}=1-\left|\int\mathrm{D} b \int\mathrm{D} b^\dagger \mathrm{e}^{\mathrm{i}S_\mathrm{eff}[b,b^\dagger]}\right|^2 =1- |\mathrm{e}^{\mathrm{i}\Gamma}|^2=1-\mathrm{e}^{-2\mathrm{Im}\Gamma},
\end{align}
where $\Gamma$ is the collection of 1 PI diagram~\cite{WenXiaoGang2007}. 
In the perturbation expansion in powers of coupling strength, the sliding-dependent contribution to $\mathrm{Im}\Gamma$ always forms a series of terms, leading to a small $\mathrm{Im}\Gamma$, i.e.,
\begin{align}
    \mathrm{e}^{-2\mathrm{Im}\Gamma} \approx 1- 2\mathrm{Im}\Gamma.
\end{align}
Thus, we the probability of particle production can be written as
\begin{align}
    \mathcal{P}=2\mathrm{Im}\Gamma.
\end{align}
After performing the functional integral, we can obtain the expression for $\Gamma$:
\begin{align}\nonumber\\&
\Gamma=\frac{\mathrm{i}}{2}\ln\det  \{\mathrm{i}\frac{\partial}{\partial t}-JS\nabla^2\nonumber\\&
+ |C(\bm{r}_{\parallel})|^2\delta(z-h)\Delta_\mathrm{R}(t-t',\bm{r}_{\parallel}-\bm{r}_{\parallel}')\delta(z'-h)\nonumber\\&
+ |C(\bm{r}_{\parallel})|^2\delta(z)\Delta_\mathrm{L}(t-t',\bm{r}_{\parallel}-\bm{r}_{\parallel}')\delta(z')\}.
\end{align}
In general, the coupling strength $|C(\bm{r}{\parallel})|$ is sufficiently small for perturbation calculations, allowing $\Gamma$ to be expanded in powers of $|C(\bm{r}{\parallel})|$.
To the order of $|C(\bm{r}_{\parallel})|^4$, $\Gamma$ can be written as
\begin{align}\nonumber\\&
\Gamma=\mathrm{const} - \mathrm{iTr}[|C|^4 G_{\rm{F}} \Delta_{\rm{R}} G_{\rm{F}} \Delta_{\rm{L}}].
\end{align}
Here the leading-order terms are not vanishing. However, physically speaking, the leading-order corresponds to the case where either the L- or R-surface is absent. Since our aim is to investigate the magnon-mediated plasmon friction between the two surfaces, we ignore the leading-order terms. Additionally, the terms in the absence of either of the two surfaces, $\mathrm{Tr}[|C|^4 G_{\rm{F}} \Delta_{\rm{R}} G_{\rm{F}} \Delta_{\rm{R}}]$ and $\mathrm{Tr}[|C|^4 G_{\rm{F}} \Delta_{\rm{L}} G_{\rm{F}} \Delta_{\rm{L}}]$, are not required, since we investigate the magnon-mediated plasmon friction between the two surfaces.
Then the lowest ordered nontrivial contribution can be expressed as follows:
\begin{align}\nonumber\\&
    \mathrm{Tr}[|C|^4 G_{\rm{F}} \Delta_{\rm{R}} G_{\rm{F}} \Delta_{\rm{L}}]\nonumber\\&
    =TA\int_{-\infty}^{\infty}\frac{d\omega}{2\pi} \int\frac{\mathrm{d}^2\bm{k}_{\parallel}}{(2\pi)^2}|C(\bm{k}_{\parallel})|^4 G_{\rm{F}}(\omega,\bm{k}_{\parallel};0-h)\Delta_{\rm{R}}(\omega,\bm{k}_{\parallel}) G_{\rm{F}}(\omega,\bm{k}_{\parallel};h-0) \Delta_{\rm{L}}(\omega,\bm{k}_{\parallel}),
\label{low_ord}
\end{align}
where $T$ is the duration of the system and $A$ is the total area of the surface. The kernels of R-plasmon and L-plasmon in frequency-momentum space can be obtained from Eq~\eqref{Eff_act_ker_R} and Eq~\eqref{Eff_act_ker_L} respectively as:
\begin{align}
    \Delta_\mathrm{R}(\omega,\bm{k}_{\parallel})
    =\frac{(\omega-\bm{v}\cdot\bm{k}_{\parallel})^2}{(\omega-\bm{v}\cdot\bm{k}_{\parallel})^2-u^2\bm{k}_{\parallel}^2-\Omega^2+\mathrm{i}\epsilon},
\label{Ker_R_plas}
\end{align}
and
\begin{align}
    \Delta_\mathrm{L}(\omega,\bm{k}_{\parallel})=\frac{\omega^2}{\omega^2-u^2\bm{k}_{\parallel}^2-\Omega^2+\mathrm{i}\epsilon}.
\label{Ker_L_plas}
\end{align}
The integrand on the complex $\omega$-plane has only two poles, which depend on the relative velocity $\bm{v}$:
\begin{align}
    \omega_\pm=\bm{v}_{\parallel} \cdot \bm{k}_{\parallel} \pm \sqrt{u^2\bm{k}_{\parallel}^2+\Omega^2} \mp \mathrm{i}\epsilon.
\end{align}
Here, we do not consider the antiparticles corresponding to the plasmons, because our model focuses on plasmon-magnon coupling, which is an effective interaction rather than a fundamental electromagnetic interaction.
After substituting Eq~\eqref{Kel_Four}, \eqref{Ker_R_plas}, and \eqref{Ker_L_plas} into Eq~\eqref{low_ord}, and evaluating the integral over $\omega$ using the residue theorem [17], we obtain the expression for $\Gamma$ as follows:
\begin{align}\nonumber\\&
    \Gamma= TA\int\frac{\mathrm{d}^2\bm{k}_{\parallel}}{(2\pi)^2}|C(\bm{k}_{\parallel})|^4 \frac{1}{\bm{v}_{\parallel} \cdot \bm{k}_{\parallel}} \frac{1}{u^2\bm{k}_{\parallel}^2+\Omega^2}\nonumber\\&
    \times\frac{JS(\bm{v}_{\parallel} \cdot \bm{k}_{\parallel} - \sqrt{u^2\bm{k}_{\parallel}^2+\Omega^2} + \mathrm{i}\epsilon)^2( - \sqrt{u^2\bm{k}_{\parallel}^2+\Omega^2} + \mathrm{i}\epsilon)^2}{\bm{v}_{\parallel} \cdot \bm{k}_{\parallel} - \sqrt{u^2\bm{k}_{\parallel}^2+\Omega^2} -JS\bm{k}_{\parallel}^2 +\mathrm{i}\epsilon}\nonumber\\&
    \times\exp(\frac{2\mathrm{i}h}{\sqrt{JS}}\sqrt{\bm{v}_{\parallel} \cdot \bm{k}_{\parallel} - \sqrt{u^2\bm{k}_{\parallel}^2+\Omega^2} -JS\bm{k}_{\parallel}^2 +\mathrm{i}\epsilon})\nonumber\\&
    \times\frac{1}{\bm{v}_{\parallel} \cdot \bm{k}_{\parallel} - 2\sqrt{u^2\bm{k}_{\parallel}^2+\Omega^2} + \mathrm{i}\epsilon}.
\label{Gamma}
\end{align}
Note that $\bm{v}_{\parallel} \cdot \bm{k}_{\parallel}$ can be regarded as the energy pumped into the system by the relative motion~\cite{volokitin2011quantumfrictiongraphene}.
Thus, by taking into account the identity
\begin{align}
    \frac{1}{a\pm\mathrm{i}\epsilon} = \mathrm{p.v.}(\frac{1}{a\pm\mathrm{i}\epsilon})\mp\mathrm{i}\pi\delta(a),
\end{align}
Eq~\eqref{Gamma} can gives two Dirac $\delta$-functions $\delta(\bm{v}_{\parallel} \cdot \bm{k}_{\parallel} - 2\sqrt{u^2\bm{k}_{\parallel}^2+\Omega^2})$ and $\delta(\bm{v}_{\parallel} \cdot \bm{k}_{\parallel} - \sqrt{u^2\bm{k}_{\parallel}^2+\Omega^2} -JS\bm{k}_{\parallel}^2)$.
The former corresponds to plasmon excitation, while the latter corresponds to magnon-plasmon hybrid excitation on the R-surface. Since the magnons only play the role in mediating, here we only focus on the former case,
and the corresponding probability of particle production can be written as
\begin{align}\nonumber\\&
    \mathcal{P}=2\mathrm{Im}\Gamma\nonumber\\&
    =2TA\int\frac{\mathrm{d}^2\bm{k}_{\parallel}}{(2\pi)^2}|C(\bm{k}_{\parallel})|^4 \frac{1}{\bm{v}_{\parallel} \cdot \bm{k}_{\parallel}} \frac{1}{u^2\bm{k}_{\parallel}^2+\Omega^2}\nonumber\\&
    \times\frac{JS(\bm{v}_{\parallel} \cdot \bm{k}_{\parallel} - \sqrt{u^2\bm{k}_{\parallel}^2+\Omega^2})^2( - \sqrt{u^2\bm{k}_{\parallel}^2+\Omega^2})^2}{\bm{v}_{\parallel} \cdot \bm{k}_{\parallel} - \sqrt{u^2\bm{k}_{\parallel}^2+\Omega^2} -JS\bm{k}_{\parallel}^2}\nonumber\\&
    \times\exp(\frac{-2h}{\sqrt{JS}}\sqrt{-\bm{v}_{\parallel} \cdot \bm{k}_{\parallel} + \sqrt{u^2\bm{k}_{\parallel}^2+\Omega^2} +JS\bm{k}_{\parallel}^2})\nonumber\\&
    \times\pi\delta(\bm{v}_{\parallel} \cdot \bm{k}_{\parallel} - 2\sqrt{u^2\bm{k}_{\parallel}^2+\Omega^2}),
\end{align}
with the condition $-\bm{v}_{\parallel} \cdot \bm{k}_{\parallel} + \sqrt{u^2\bm{k}_{\parallel}^2+\Omega^2} +JS\bm{k}_{\parallel}^2>0$, which means no magnon being excited. Without loss of generality, we set the velocity $\bm{v}_{\parallel}$ along the $x$-direction, i.e.,
\begin{align}
\bm{v}_{\parallel}=\begin{pmatrix}v,0\end{pmatrix}.
\label{Velocity}
\end{align}
Thus, the probability of particle production becomes
\begin{align}\nonumber\\&
    \mathcal{P}=2\mathrm{Im}\Gamma\nonumber\\&
    =2TA\int\frac{\mathrm{d}^2\bm{k}_{\parallel}}{(2\pi)^2}|C(\bm{k}_{\parallel})|^4 \frac{1}{vk_x} \frac{1}{u^2\bm{k}_{\parallel}^2+\Omega^2}\nonumber\\&
    \times\frac{JS\left\{vk_x - \sqrt{u^2\bm{k}_{\parallel}^2+\Omega^2}\right\}^2\left\{ - \sqrt{u^2\bm{k}_{\parallel}^2+\Omega^2}\right\}^2}{vk_x - \sqrt{u^2\bm{k}_{\parallel}^2+\Omega^2} -JS\bm{k}_{\parallel}^2}\nonumber\\&
    \times\exp\left\{\frac{-2h}{\sqrt{JS}}\sqrt{-vk_x + \sqrt{u^2\bm{k}_{\parallel}^2+\Omega^2} +JS\bm{k}_{\parallel}^2}\right\}\nonumber\\&
    \times\pi\delta\left\{vk_x - 2\sqrt{u^2\bm{k}_{\parallel}^2+\Omega^2}\right\},
 \label{P_p_p}
\end{align}
By using the property of Dirac $\delta$-function
\begin{align}
    \delta[f(x)]=\sum_i \left\|\frac{\mathrm{d}f(x_i)}{\mathrm{d}x_i}\right\|^{-1}\delta(x-x_i),
\end{align}
the item of Dirac $\delta$-function in Eq~\eqref{P_p_p} can be expressed as
\begin{align}
    \delta\left\{vk_x - 2\sqrt{u^2\bm{k}_{\parallel}^2+\Omega^2}\right\}=\frac{v}{v^2-4u^2}
    \left\{\delta\left\{ k_x-2\sqrt{\frac{u^2k_y^2+\Omega^2}{v^2-4u^2}}\right\}
    +\delta\left\{ k_x+2\sqrt{\frac{u^2k_y^2+\Omega^2}{v^2-4u^2}}\right\}\right\},
\end{align}
which gives a threshold speed $v_P \geq 2u$, i.e., the plasmons are excited when the relative motion speed is bigger than $2u$. This threshold speed is similar to the quantum friction between two graphene sheets~\cite{PhysRevD.95.065012}. Obviously, here the probability decreases as the thickness of the ferromagnet increases.

\begin{figure}[htbp]
  \centering
  \includegraphics[width=5in]{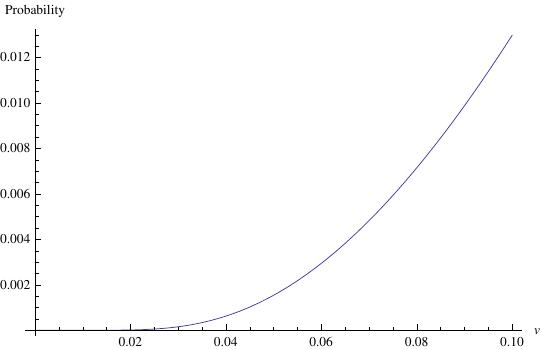}\\
  \caption{The probability of particle production as a function of $v$ in units of $TAC^4$, with parameters $\Omega=0.3,JS=0.6$ and $h=0.1$.}
  \label{fig_pp}
\end{figure}

To observe how the probability changes with increasing velocity $v$, we set the coupling strength $C(\bm{k}_{\parallel})$ to a constant value, denoted by the real-valued parameter $C$, and temporarily ignore the dispersion of the plasmons by setting $u=0$. Then the probability of particle production can be expressed as follows
\begin{align}
    \mathcal{P}=TA\int_{-\infty}^\infty\frac{\mathrm{d}k_y}{2\pi}
    \frac{2\pi C^4}{2\Omega}\frac{JS\Omega^2}{\Omega-JS[(2\Omega/v)^2+k_y^2]}
    \mathrm{e}^{-\frac{2h}{\sqrt{JS}} \sqrt{-\Omega+JS[(2\Omega/v)^2+k_y^2]}}.
\end{align}
Here we calculate the integral over $k_y$ numerically and show the result in Figure~\ref{fig_pp}, which indicates a very significant and positive correlation between probability $\mathcal{P}$ and speed $v$.  In addition, the threshold speed $v_\mathrm{P}$ tend to zero naturally due to the elimination of the plasmons' speed $u$.

\section{Frictional force}\label{friction}
\begin{figure}[htbp]
  \centering
  \includegraphics[width=5in]{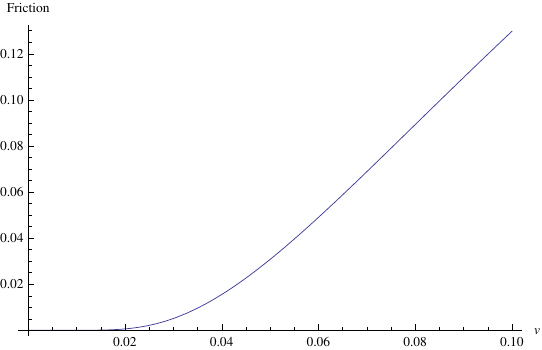}\\
  \caption{The frictional force as a function of $v$ in units of $TAC^4$, with parameters $\Omega=0.3,JS=0.6$ and $h=0.1$.}\label{fig_fric}
\end{figure}

According to the physical significance of $\Gamma$, the energy accumulated by the excitations due to the relative motion can be expressed as:
\begin{align}\nonumber\\&
    \mathcal{E}=TA\int_{-\infty}^\infty\frac{\mathrm{d}\omega}{2\pi} \mathrm{Im}\Big\{\int\frac{\mathrm{d}^2\bm{k}_{\parallel}}{(2\pi)^2} |\omega||C(\bm{k}_{\parallel})|^4 \nonumber\\&
    \times G_{\rm{F}}(\omega,\bm{k}_{\parallel};0-h)\Delta_{\rm{R}}(\omega,\bm{k}_{\parallel}) G_{\rm{F}}(\omega,\bm{k}_{\parallel};h-0) \Delta_{\rm{L}}(\omega,\bm{k}_{\parallel})\Big\},
\end{align}
And then we can write the dissipation power per unit area as
\begin{align}\nonumber\\&
    \bm{F}_\mathrm{fric}\cdot\bm{v}_{\parallel}=\frac{\mathcal{E}}{TA}
    =2\int\frac{\mathrm{d}^2\bm{k}_{\parallel}}{(2\pi)^2}|C(\bm{k}_{\parallel})|^4 \frac{1}{\bm{v}_{\parallel} \cdot \bm{k}_{\parallel}} \frac{1}{u^2\bm{k}_{\parallel}^2+\Omega^2}\nonumber\\&
    \times\frac{JS\left\{\bm{v}_{\parallel} \cdot \bm{k}_{\parallel} - \sqrt{u^2\bm{k}_{\parallel}^2+\Omega^2}\right\}^3\left\{ - \sqrt{u^2\bm{k}_{\parallel}^2+\Omega^2}\right\}^2}{\bm{v}_{\parallel} \cdot \bm{k}_{\parallel} - \sqrt{u^2\bm{k}_{\parallel}^2+\Omega^2} -JS\bm{k}_{\parallel}^2}\nonumber\\&
    \times\exp\left\{\frac{-2h}{\sqrt{JS}}\sqrt{-\bm{v}_{\parallel} \cdot \bm{k}_{\parallel} + \sqrt{u^2\bm{k}_{\parallel}^2+\Omega^2} +JS\bm{k}_{\parallel}^2}\right\}\nonumber\\&
    \times\pi\delta\left\{\bm{v}_{\parallel} \cdot \bm{k}_{\parallel} - 2\sqrt{u^2\bm{k}_{\parallel}^2+\Omega^2}\right\},
\end{align}
where $\bm{F}_\mathrm{fric}$ is the friction acting on the R-surface. After taking the same setup of Eq~\eqref{Velocity} as $\mathrm{Im}\Gamma$, the frictional force becomes:
\begin{align}\nonumber\\&
    F_\mathrm{fric}
    =2\int\frac{\mathrm{d}^2\bm{k}_{\parallel}}{(2\pi)^2}|C(\bm{k}_{\parallel})|^4 \frac{1}{v}\frac{1}{vk_x} \frac{1}{u^2\bm{k}_{\parallel}^2+\Omega^2}\nonumber\\&
    \times\frac{JS(vk_x - \sqrt{u^2\bm{k}_{\parallel}^2+\Omega^2})^3( - \sqrt{u^2\bm{k}_{\parallel}^2+\Omega^2})^2}{vk_x - \sqrt{u^2\bm{k}_{\parallel}^2+\Omega^2} -JS\bm{k}_{\parallel}^2}\nonumber\\&
    \times\exp\left\{\frac{-2h}{\sqrt{JS}}\sqrt{-vk_x + \sqrt{u^2\bm{k}_{\parallel}^2+\Omega^2} +JS\bm{k}_{\parallel}^2}\right\}\nonumber\\&
    \times\pi\delta \left\{ vk_x - 2\sqrt{u^2\bm{k}_{\parallel}^2+\Omega^2}\right\}.
\end{align}
Like the particles production rate of Eq~\eqref{P_p_p}, it has the same Dirac $\delta$-function, thus, the frictional force also corresponds to the same threshold speed $v_\mathrm{P}=2u$. Similar to $\mathcal{P}$, the frictional force can be expressed as
\begin{align}
    F_\mathrm{fric}=\int_{-\infty}^\infty\frac{\mathrm{d}k_y}{2\pi}
    \frac{2\pi C^4}{2v\Omega}\frac{JS\Omega^3}{\Omega-JS[(2\Omega/v)^2+k_y^2]}
    \mathrm{e}^{-\frac{2h}{\sqrt{JS}} \sqrt{-\Omega+JS[(2\Omega/v)^2+k_y^2]}}.
\end{align}
The numerical result of Eq (41) is shown in Figure \ref{fig_fric}, and like most frictional forces, here it is also positively correlated with speed.

\section{Conclusion}\label{conclusion}
In this paper we have studied a magnon mediated friction between two metallic surfaces separated by a ferromagnetic intermedium with thickness $h$. The internal DOFs in the two metallic surfaces are assumed to be plasmons, while the excitations in the intermediate material are magnons. To study the frictional force in the system, we assume the R-surface is moving uniformly parallel to the L-surface at a constant velocity, creating relative sliding. The derivation starts from the effective action of the magnons, and then we analyse the free Green's function of the magnons since these magnons carry the interaction between the two surfaces. To handle the energy transfer in the system for friction, it is necessary to address the probability of particle production.The theoretical approach to particle production probability shows that a threshold speed exists, and numerical analysis indicates a positive correlation between the probability and the speed. Finally, we derived the frictional force of the system, both the theoretical and numerical results agree with the behaviors of the particle production probability, i.e., the friction also has a positive correlation with the moving speed.

Furthermore, it is interesting to consider the case of a system with magnon-plasmon hybrid excitation. The probability of particle production can be expressed as
\begin{align}\nonumber\\&
    \mathcal{P}=2\int\frac{\mathrm{d}^2\bm{k}_{\parallel}}{(2\pi)^2}|C(\bm{k}_{\parallel})|^4 \frac{1}{\bm{v}_{\parallel} \cdot \bm{k}_{\parallel}} \frac{1}{u^2\bm{k}_{\parallel}^2+\Omega^2}\nonumber\\&
    \times JS(\bm{v}_{\parallel} \cdot \bm{k}_{\parallel} - \sqrt{u^2\bm{k}_{\parallel}^2+\Omega^2})^2( - \sqrt{u^2\bm{k}_{\parallel}^2+\Omega^2} )^2\nonumber\\&
    \times\frac{1}{\bm{v}_{\parallel} \cdot \bm{k}_{\parallel} - 2\sqrt{u^2\bm{k}_{\parallel}^2+\Omega^2}}\pi\delta(\bm{v}_{\parallel} \cdot \bm{k}_{\parallel} - \sqrt{u^2\bm{k}_{\parallel}^2+\Omega^2} -JS\bm{k}_{\parallel}^2),
\end{align}
with the condition $-\bm{v}_{\parallel} \cdot \bm{k}_{\parallel} + \sqrt{u^2\bm{k}_{\parallel}^2+\Omega^2} +JS\bm{k}_{\parallel}^2<0$. Comparing Eq (40) with Eq (30), we can see that it is no exponential factor in Eq (40), which means the probability of magnon-plasmon hybrid excitations does not decrease as the thickness of the ferromagnet increases. This indicates that the corresponding frictional force must be independent of $h$.

\section*{Acknowledgement}
We gratefully acknowledge the fruitful discussions with Qiang Sun.
This research was supported by Dongying Science Development Fund(No. DJB2023015).

\vspace*{2mm}

\printbibliography

@article{Spencer1999,
    author = {Persson, Bo N. J. and Spencer, Nicholas D.},
    title = "{Sliding Friction: Physical Principles and Applications}",
    journal = {Physics Today},
    volume = {52},
    number = {1},
    pages = {66-68},
    year = {1999},
    month = {01},
    issn = {0031-9228},
    doi = {10.1063/1.882557}
}

@article{PERSSON199983,
title = {Sliding friction},
journal = {Surface Science Reports},
volume = {33},
number = {3},
pages = {83-119},
year = {1999},
issn = {0167-5729},
doi = {https://doi.org/10.1016/S0167-5729(98)00009-0},
author = {B.N.J. Persson}
}

@article{Mason2001,
author = {Mason, B. and Winder, S. and Krim, Jacqueline},
year = {2001},
month = {01},
pages = {59-65},
title = {On the current status of quartz crystal microbalance studies of superconductivity-dependent sliding friction},
volume = {10},
journal = {Tribology Letters - TRIBOL LETT},
doi = {10.1023/A:1009042816366}
}

@article{PhysRevLett.80.1690,
  title = {Superconductivity-Dependent Sliding Friction},
  author = {Dayo, A. and Alnasrallah, W. and Krim, J.},
  journal = {Phys. Rev. Lett.},
  volume = {80},
  issue = {8},
  pages = {1690--1693},
  numpages = {0},
  year = {1998},
  month = {Feb},
  publisher = {American Physical Society},
  doi = {10.1103/PhysRevLett.80.1690}
}

@article{Pendry1997,
doi = {10.1088/0953-8984/9/47/001},
year = {1997},
month = {nov},
publisher = {},
volume = {9},
number = {47},
pages = {10301},
author = {J B Pendry},
title = {Shearing the vacuum - quantum friction},
journal = {Journal of Physics: Condensed Matter}
}

@article{PhysRevB.72.224101,
  title = {Quantum friction in nanomechanical oscillators at millikelvin temperatures},
  author = {Zolfagharkhani, Guiti and Gaidarzhy, Alexei and Shim, Seung-Bo and Badzey, Robert L. and Mohanty, Pritiraj},
  journal = {Phys. Rev. B},
  volume = {72},
  issue = {22},
  pages = {224101},
  numpages = {5},
  year = {2005},
  month = {Dec},
  publisher = {American Physical Society},
  doi = {10.1103/PhysRevB.72.224101}
}

@article{Kheiri2023,
author = {Kheiri, Rasoul},
journal = {Friction},
year = {2023},
month = {05},
volume = {11},
pages = {1877–1894},
title = {Quantum “contact” friction: The contribution of kinetic friction coefficient from thermal fluctuations},
doi = {https://doi.org/10.1007/s40544-022-0719-1}
}

@article{Wangyang2022,
author = {Wang, Yang and Jia, Yu},
title = {Quantum dissipation and friction attributed to plasmons},
journal = {Modern Physics Letters B},
volume = {36},
number = {06},
pages = {2150589},
year = {2022},
doi = {10.1142/S0217984921505898}
}

@article{PhysRevB.108.045406,
  title = {Negative vacuum friction in terahertz gain systems},
  author = {Ge, Lixin},
  journal = {Phys. Rev. B},
  volume = {108},
  issue = {4},
  pages = {045406},
  numpages = {6},
  year = {2023},
  month = {Jul},
  publisher = {American Physical Society},
  doi = {10.1103/PhysRevB.108.045406}
}

@article{Khosravi2024,
doi = {10.1088/1367-2630/ad3fe1},
year = {2024},
month = {05},
publisher = {IOP Publishing},
volume = {26},
number = {5},
pages = {053006},
author = {Farhad Khosravi and Wenbo Sun and Chinmay Khandekar and Tongcang Li and Zubin Jacob},
title = {Giant enhancement of vacuum friction in spinning YIG nanospheres},
journal = {New Journal of Physics}
}

@article{PhysRevLett.101.137205,
  title = {Magnetic Friction in Ising Spin Systems},
  author = {Kadau, Dirk and Hucht, Alfred and Wolf, Dietrich E.},
  journal = {Phys. Rev. Lett.},
  volume = {101},
  issue = {13},
  pages = {137205},
  numpages = {4},
  year = {2008},
  month = {Sep},
  publisher = {American Physical Society},
  doi = {10.1103/PhysRevLett.101.137205}
}

@article{Miura2012,
author = {Miura ,Daisuke and Sakuma ,Akimasa},
title = {Microscopic Theory of Magnon-Drag Thermoelectric Transport in Ferromagnetic Metals},
journal = {Journal of the Physical Society of Japan},
volume = {81},
number = {11},
pages = {113602},
year = {2012},
doi = {10.1143/JPSJ.81.113602}
}

@ARTICLE{Magiera5257437,
  author={Magiera, Martin P. and Wolf, Dietrich E. and Brendel, Lothar and Nowak, Ulrich},
  journal={IEEE Transactions on Magnetics}, 
  title={Magnetic Friction and the Role of Temperature}, 
  year={2009},
  volume={45},
  number={10},
  pages={3938-3941},
  doi={10.1109/TMAG.2009.2023623}
}

@article{Wang2022wyh,
    author = "Wang, Yang and Jia, Yu",
    title = "{Dissipation and friction of a quantum spin system}",
    doi = "10.1140/epjb/s10051-022-00330-z",
    journal = "Eur. Phys. J. B",
    volume = "95",
    number = "4",
    pages = "75",
    year = "2022"
}

@article{PhysRevA.99.042511,
  title = {Casimir friction between a magnetic and a dielectric material in the nonretarded limit},
  author = {H\o{}ye, Johan S. and Brevik, Iver},
  journal = {Phys. Rev. A},
  volume = {99},
  issue = {4},
  pages = {042511},
  numpages = {8},
  year = {2019},
  month = {Apr},
  publisher = {American Physical Society},
  doi = {10.1103/PhysRevA.99.042511}
}

@article{Costa2023,
author = {Costa, Antonio and Vasilevskiy, Mikhail and Fernández-Rossier, J. and Peres, Nuno},
year = {2023},
month = {05},
pages = {},
title = {Strongly Coupled Magnon-Plasmon Polaritons in Graphene-Two-Dimensional Ferromagnet Heterostructures},
volume = {23},
journal = {Nano letters},
doi = {10.1021/acs.nanolett.3c00907}
}

@article{PhysRevB.108.045414,
  title = {Magnon-plasmon hybridization mediated by spin-orbit interaction in magnetic materials},
  author = {Dyrda\l{}, Anna and Qaiumzadeh, Alireza and Brataas, Arne and Barna\ifmmode \acute{s}\else \'{s}\fi{}, J\'ozef},
  journal = {Phys. Rev. B},
  volume = {108},
  issue = {4},
  pages = {045414},
  numpages = {6},
  year = {2023},
  month = {Jul},
  publisher = {American Physical Society},
  doi = {10.1103/PhysRevB.108.045414}
}

@article{ERastelli1979,
doi = {10.1088/0022-3719/12/10/021},
url = {https://dx.doi.org/10.1088/0022-3719/12/10/021},
year = {1979},
month = {may},
publisher = {},
volume = {12},
number = {10},
pages = {1899},
author = {E Rastelli and  P -A Lindgard},
title = {Exact results for spin-wave renormalisation in Heisenberg, and planar ferromagnets},
journal = {Journal of Physics C: Solid State Physics}
}

@book{WenXiaoGang2007,
    author = {Wen, Xiao-Gang},
    title = "{Quantum Field Theory of Many-Body Systems: From the Origin of Sound to an Origin of Light and Electrons}",
    publisher = {Oxford University Press},
    year = {2007},
    month = {09},
    isbn = {9780199227259},
    doi = {10.1093/acprof:oso/9780199227259.001.0001}
}

@misc{volokitin2011quantumfrictiongraphene,
      title={Quantum friction and graphene}, 
      author={Aleksandr Volokitin},
      year={2011},
      eprint={1112.4912},
      archivePrefix={arXiv}
}

@article{PhysRevD.95.065012,
  title = {Quantum friction between graphene sheets},
  author = {Farias, M. Bel\'en and Fosco, C\'esar D. and Lombardo, Fernando C. and Mazzitelli, Francisco D.},
  journal = {Phys. Rev. D},
  volume = {95},
  issue = {6},
  pages = {065012},
  numpages = {12},
  year = {2017},
  month = {Mar},
  publisher = {American Physical Society},
  doi = {10.1103/PhysRevD.95.065012}
}

@article{PhysRevB.77.174426,
  title = {Magnetic friction of a nanometer-sized tip scanning a magnetic surface: Dynamics of a classical spin system with direct exchange and dipolar interactions between the spins},
  author = {Fusco, C. and Wolf, D. E. and Nowak, U.},
  journal = {Phys. Rev. B},
  volume = {77},
  issue = {17},
  pages = {174426},
  numpages = {5},
  year = {2008},
  month = {May},
  publisher = {American Physical Society},
  doi = {10.1103/PhysRevB.77.174426}
}

%\end{CJK*}

\end{document}